\documentclass{aa}
\usepackage{graphicx}
\usepackage{times}
\usepackage{latexsym}
\usepackage{epsfig}
\usepackage{txfonts}
\usepackage{natbib}
\usepackage{supertabular,multirow,longtable,rotating}
\usepackage{amssymb}
\usepackage{color}

\def\kms{km\,s$^{-1}$}

\def\ie{\emph{i.e.}}

\bibpunct{(}{)}{;}{a}{}{,} 

\begin{document}

   \title{On the polarisation of the\\ Red Rectangle optical emission bands~\thanks{%
   Based on observations obtained at the Canada-France-Hawaii Telescope (CFHT) which is 
   operated by the National Research Council of Canada, the Institut National des Sciences de l'Univers
   of the Centre National de la Recherche Scientique of France, and the University of Hawaii.}}

   \author{N.~L.~J.~Cox \inst{1}
   	\and
   	   B.~H.~Foing  \inst{2} 
	\and
	   J.~Cami \inst{3,5}
	\and
	   P.~J.~Sarre \inst{4}
	      }

   \institute{Instituut voor Sterrenkunde, K.U. Leuven, Celestijnenlaan 200D, 3001 Leuven, Belgium
	     \and
	     ESA Research and Scientific Support Department, P.O. Box 299, 2200-AG~~Noordwijk, The Netherlands
             \and
	     Department of Physics \& Astronomy, University of Western Ontario, London, ON N6A 3K7, Canada 
	     \and
	     School of Chemistry, The University of Nottingham, University Park, Nottingham, NG7 2RD, United Kingdom
	     \and
	     SETI Institute, 189 Bernardo Ave., Suite 100, Mountain View, CA 94043, United States
	      }

   \offprints{N.L.J.~Cox, \email{nick@ster.kuleuven.be}}

   \date{}

   \abstract{
   The origin of the narrow optical emission bands seen toward the Red Rectangle is not yet understood. 
   In this paper we investigate further the proposal that these are due to luminescence of large carbonaceous molecules.
   }{%
   The aim is to measure the polarisation of the optical narrow Red Rectangle bands (RRBs). 
   Polarised signals of several percent could be expected from certain asymmetric molecular rotators.
   }{%
   The ESPaDOnS \'{e}chelle spectrograph mounted at the CFHT was used to obtain high-resolution optical spectropolarimetric 
   data of the Red Rectangle nebular emission.
   }{%
   The RRBs at 5800, 5850, and 6615~\AA\ are detected in spectra of the nebular emission 7\arcsec\ and 13\arcsec\ North-East 
   from the central star. The 5826~\AA\ and 6635~\AA\ RRB are detected only at the position nearest to the central star.
   For both positions the Stokes $Q$ and $U$ spectra show no unambiguous polarisation signal in any of the RRBs. 
   We derive an upper limit of 0.02\% line polarisation for these RRBs. A tentative feature with peak polarisation of 0.05\% is seen
   for the 5800~\AA\ RRB at 7\arcsec\ offset. However, the Null spectra suggest that this may be an instrumental artifact.
   }{%
   The lack of a clear polarisation signal for the five detected RRBs implies that, if the emission is due to luminescence
   of complex organics, these gas-phase molecular carriers are likely to have a high degree of symmetry, as they do not exhibit a Q-branch
   in their rotational profile, although this may be modified by statistical effects.
   }


   \titlerunning{On the polarisation of the Red Rectangle optical emission bands}
   \authorrunning{Cox et al.}

   \maketitle


\section{Introduction}\label{sec:intro}
The Red Rectangle (RR) is a unique bi-conical reflection nebula (\citealt{2004AJ....127.2362C}). 
At its centre lies a close binary system of which the primary (HD~44179) is a luminous evolved star. 
The system is visually obscured by a 
thick and massive circumbinary dusty disk that is seen nearly edge-on (\citealt{1998Natur.391..868W}). 
Interactions between the stellar wind and disk result in the X-shaped nebula.
The central star is of 9th visual magnitude, while the nebula has a surface $V$ brightness of about 
19~magnitudes per square arcsecond (\emph{e.g.} \citealt{1991ApJ...383..698S}).

The infrared (IR) spectrum of the RR nebula shows strong infrared emission bands across the near- and mid-infrared spectral region.
(\citealt{1977ivsw.conf..446M}, \citealt{1977ApJ...213...66R,1978ApJ...220..568R}, \citealt{1989ApJ...341..278G},
\citealt{1996A&A...315L.245W}, \citealt{2001A&A...370.1030H}, \citealt{2002A&A...390.1089P}, \citealt{2007MNRAS.380..979S}). 
These are generally accepted to be due to vibrational transitions of PAHs (see review by \citealt{2008ARA&A..46..289T}).
Observations show that the aromatic emission in the RR originates from the extended nebular emission of the RR,
similar to the RRBs (\citealt{1993ApJ...409..412S}, \citealt{1998Natur.391..868W}).

At optical wavelengths, the RR was the first object in which Extended Red Emission (ERE) was detected (\citealt{1975ApJ...196..179C}, 
\citealt{1977PASP...89..131G}, \citealt{1980ApJ...239L.133S}, \citealt{1991ApJ...383..698S}). 
The carriers of this broad emission in the red part of the spectrum have not been identified, but it is believed that 
ERE is due to photoluminescence from materials such as PAHs, quenched carbonaceous composites, 
hydrogenated amorphous carbon, carbon clusters, nanodiamonds etc. (\citealt{1986A&A...170...91D}, 
\citealt{1990ApJ...355..182W}, \citealt{1990ApJ...364L..45F}, \citealt{2006ApJ...636..303W}). 
Note that the ERE has also been detected in the diffuse ISM (\citealt{1998ApJ...498..522G})
and in compact \ion{H}{ii} regions (\citealt{2000A&A...364..723D}).
In addition, blue luminescence (BL) between about 3900 and 4100~\AA\ has also been detected toward both the RR and several reflection nebulae, and has been
attributed specifically to fluorescence of (neutral) PAHs containing 14 to 18 carbon atoms (\citealt{2004ApJ...606L..65V,2005ApJ...633..262V}).
Recent laboratory experiments suggest that BL is due to small gas-phase PAHs while ERE is due to larger PAHs in grains (\citealt{2009ApJ...690..111W}).

Superposed on the ERE in the Red Rectangle are a plethora of narrow unidentified emission features (\citealt{1980ApJ...239L.133S},
\citealt{1981Natur.292..317W}),  whose widths, profile shapes and  peak positions change as a function of the distance from
the central star (\citealt{1991ApJ...383..698S}, \citealt{1992MNRAS.255P..11S}, \citealt{1995Sci...269..674S}, \citealt{2002MNRAS.332L..17G},
\citealt{2002A&A...390..147V}) and are called the Red Rectangle bands (RRBs). The ERE
and the RRBs are most pronounced along the bi-conical interfaces although their spatial distributions are not identical
(\citealt{1991ApJ...383..698S}). 
As one moves away from the central star along the bi-conical interface, the wavelengths of 
the RRBs shift bluewards (\citealt{2002MNRAS.332L..17G}, \citealt{2002A&A...390..147V}).

\citet{1980ApJ...239L.133S} suggested that the sharp emission features (\emph{e.g.} near 5800 and 6615~\AA) may arise in gaseous molecules, 
whereas the diffuse underlying (ERE) component would correspond to the same transition (but somewhat perturbed) in molecules attached to grains.
The RRBs are nebular emission rather than scattered starlight.
\citet{2006ApJ...639..194S} illustrate that the RRBs can be reproduced, for example, by emission due to (vibrational) ``sequence structure'' 
of large fluorescent PAH molecules, where overlapping vibrational contours at different vibrational temperatures cause the 
observed variations across the nebula. 
To the best of our knowledge the polarisation properties of the bands have not been investigated 
prior to the work reported here.
A recent assesment of the Red Rectangle band problem is given in \citet{2009Ap&SS.323..337G} who argue that some aspects of the 
RRB behaviour may be be due to 1) self-absorption (cold material in front of hotter material), 2) velocity dispersion within 
the nebula (broadening of intrinsic line profile), or 3) overlapping of bands from different weak emitters.

A key issue for the Red Rectangle emission bands (and diffuse interstellar absorption bands) 
is whether the spectral line carriers are free gas phase molecules or are directly associated with dust grains. 
This latter scenario could hold if the carriers were within or on the surface of the grains.
Interstellar grains cause polarisation of starlight and so polarimetric measurements on the Red Rectangle bands allows this to be probed.

In addition to the intrinsic interest in understanding and assignment of the RRBs,
some of the prominent RRBs at 5800, 6380 and 6615~\AA\ are very close in wavelength (where the differences decreases with increasing distance 
from the central star) to a set of strong narrow, but diffuse, interstellar absorption lines at 5797, 6379 and 6614~\AA.
This coincidence has led to the hypothesis that there may be a direct link between the RRBs and some of 
the ubiquitous diffuse interstellar bands (\citealt{1991Natur.353..393F}, \citealt{1991Natur.351..356S}, \citealt{1992MNRAS.255P..11S}, \citealt{1995Sci...269..674S}).
Diffuse band carriers are also believed to be large, stable carbonaceous interstellar gas-phase molecules (\citealt{2006JMoSp.238....1S}) and
any connection between DIBs and RRBs could therefore provide further insight in either carrier.
For now both their identity and connection to the RRB are outstanding issues.

Broad-band spectropolarimetry 11\arcsec\ south of the central star revealed a reduction in the degree of polarization (from 6\% in the
blue down to 2\% at the ERE wavelength) but constant position angle ($\sim$78\degr). This suggests that the emission originates within the
nebula itself and dilutes the polarised scattered light from the central star \citep{1980ApJ...239L.133S,2009MNRAS.392.1217G}. 
Substantial polarisation ($\sim$20\%) is observed along the conic interface just outside the bicone, while values of 5-10\% are common 
in the nebula itself, although the ERE component is likely unpolarised (\citealt{1981MNRAS.196..635P}).
\citet{1996ApJ...467L.105R} measured a linear continuum polarisation level of 2.2$\pm$1.1\% toward the central star.

From theoretical work and laboratory experiments it has been found that fluorescent emission of large molecules 
can exhibit linear polarisation changes along a set of vibronic lines (\citealt{1986OptSp..61...48T,1995SpecL..28..441T,
1998JApSp..65..883T,1998JApSp..65..643T,1998OptSp..84..839T,2002JApSp..69..711T,
2002JApSp..69..571B,2006DokPh..51...17B,2009JApSp..76..806T}). 
The overall degree of polarization as well as the variation across the rotational contour depends strongly on the molecular symmetry 
(see also \citealt{1986JChPh..85.4311N}): extreme prolate or oblate molecules could display polarizations of up to 20\%.
Very symmetrical species on the other hand (or indeed dust particles) are not expected to exhibit much polarization across the lines. 
\citet{2009ApJ...698.1292S} showed that the infrared emission of PAHs could become polarised, albeit at low levels of $<$0.1\%, 
upon anisotropic illumination by UV photons.
This offers tantalizing prospects for constraining the geometry (shape) of the RRB carriers by measuring the linear line 
polarization in space for the first time.

This paper presents a spectropolarimetric study of the nebulosity of the Red Rectangle. 
After a presentation of the observations we discuss the implications of our results for the properties of the carriers of the 
optical narrow emission bands (RRBs) and, by implication, possibly also those of the DIB and UIB car\-riers.

\begin{figure}[t!]
 \centering
  \includegraphics[angle=0,width=\columnwidth,clip]{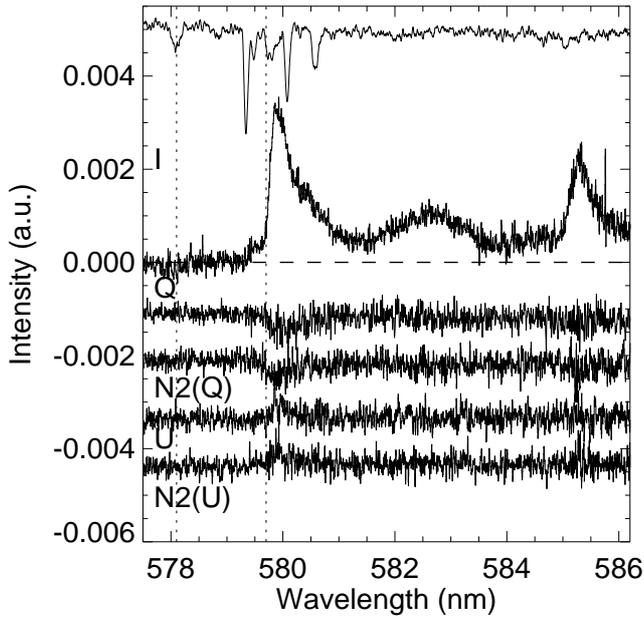}
  \includegraphics[angle=0,width=\columnwidth,clip]{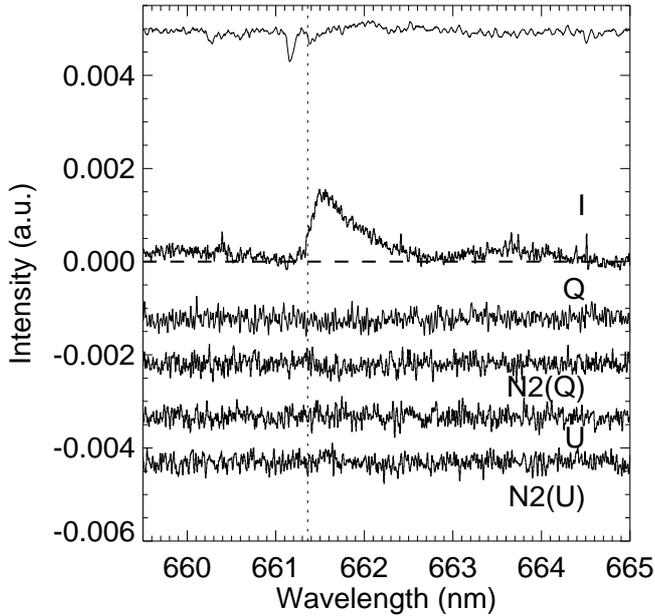}
   \caption{Stokes I, Q and U spectra of the 5800 and 5826~\AA\ (top), 
   and 6615~\AA\ (bottom) RRBs at an offset of 7\arcsec.
   For the displayed spectra a 3-pixel smoothing function was applied.   
   The spectrum (continuum normalised and arbitrarily scaled) at the top is that of the central star.}
   \label{fig:rrb5}
\end{figure}

\begin{figure}[t!]
 \centering
  \includegraphics[angle=0,width=\columnwidth,clip]{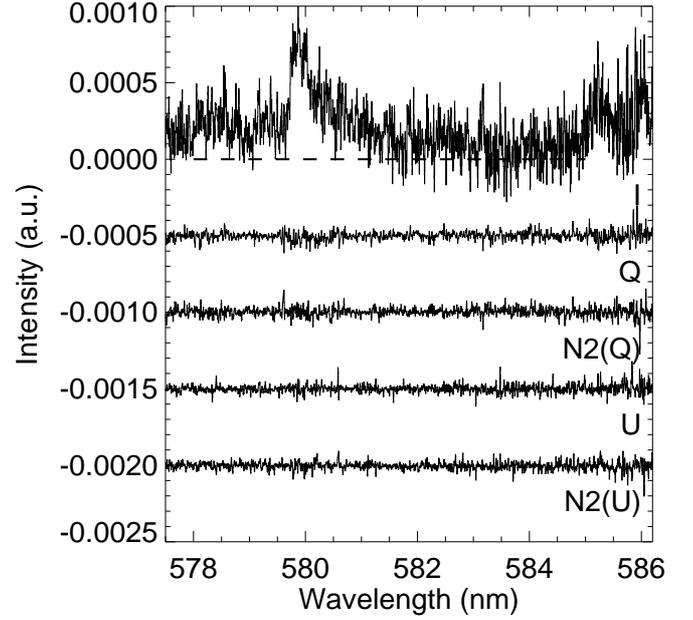}
  \includegraphics[angle=0,width=\columnwidth,clip]{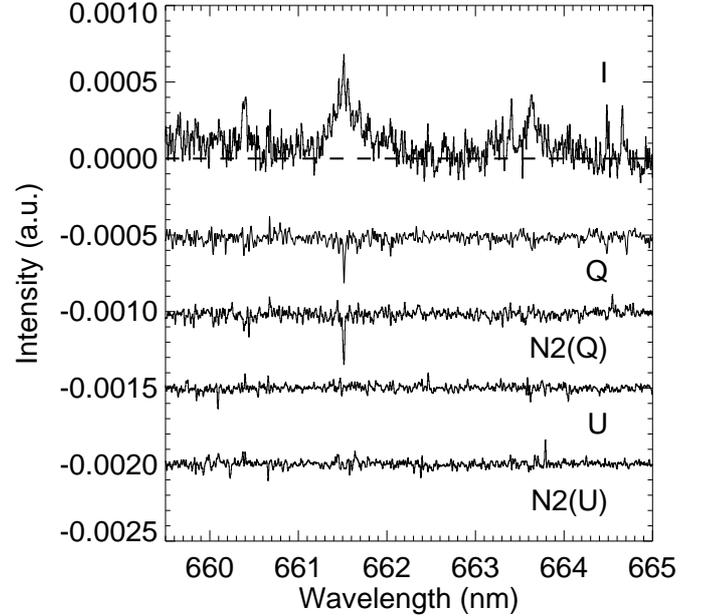}
   \caption{Stokes I, Q and U spectra of the 5800~\AA\ (top) and 6616~\AA\ (bottom) RRBs at an 
   offset of 13\arcsec. 
   For the displayed spectra a 3-pixel smoothing function was applied.
   In the average intensity spectrum a narrow feature appears at the location 
   of the broad 6635~\AA\ RRB. The very sharp peaks are probably instrumental artifacts.} 
   \label{fig:rrb10}
\end{figure}

\section{Spectropolarimetric observations}\label{sec:observations}
For the present study we obtained new spectropolarimetry data with ESPaDOnS at the Canadian-French-Hawaian Telescope (CFHT). 
The data were taken, in service mode, during several nights in January and February 2010.
ESPaDOnS is a high-resolution high-efficiency 2-fiber echelle spectrograph with polarising capabilities. 
The resolution for the spectropolarimeter is about 64\,000,
covering a wide spectral range, from 3700 to 10480~\AA\ with only a few small gaps in the near-infrared.
The fiber (1.6\arcsec\ diameter) was placed first 4.5\arcsec\ North, 5\arcsec\ East and secondly 9\arcsec\ North, 9\arcsec\ East,
along the bi-conical interface, relative to central star. 
In the following we refer to these two nebular positions as $\sim$7\arcsec\ and $\sim$13\arcsec\ offset from the central star, along
the biconical interface in the North-East direction.
For each Stokes parameter the observing sequence consists of four spectra obtained at different orientations of the prism.
For both linear Stokes parameters (\emph{i.e.} $Q$ and $U$) we obtain 3 sequences of 4x1000s at the 7\arcsec\ position (\ie\ 12\,000 seconds total exposure) and 4 sequences of 4x1200s at the 13\arcsec\ position (19\,200 seconds total).
In addition, for reference, we obtained in similar fashion Stokes $Q$ and $U$ spectra of the central star, HD~44179 (5\,040 seconds total).

The observations were automatically reduced with Upena, which is the CFHT's reduction pipeline for ESPaDOnS. 
The Upena data reduction system uses Libre-ESpRIT which is a purpose built (proprietary) data reduction software tool (\citealt{1997MNRAS.291..658D}).
Due to the low flux levels of the extended emission the automatic continuum normalisation did not perform well and was therefore not used.
Absolute flux calibration was not performed. The relative spectral response was obtained via normal flat-fielding.
At the selected positions the nebula has a low surface brightness and only a weak spectral signal is received.
The heliocentric velocity correction and the radial velocity correction from telluric lines are automatically applied.
Next, the individual spectra were rebinned (to a constant wavelength grid with a pixel size of 0.03~\AA\ at 5800~\AA\ and to 0.04~\AA\ at 6600~\AA) 
and cosmic spikes removed. The spectra were then averaged with weights taken from the uncertainties given for the total intensity.
The averaged spectra at 7\arcsec\ reveal the RRBs peaking at 5799, 5826, 5854, 6615, and 6635~\AA.
Additional weaker RRBs have been revealed in previous studies by \emph{e.g.} \cite{2002A&A...390..147V},
but these do not show up clearly in our spectra. 

\begin{figure}[t!]
 \centering
  \includegraphics[angle=0,width=\columnwidth,clip]{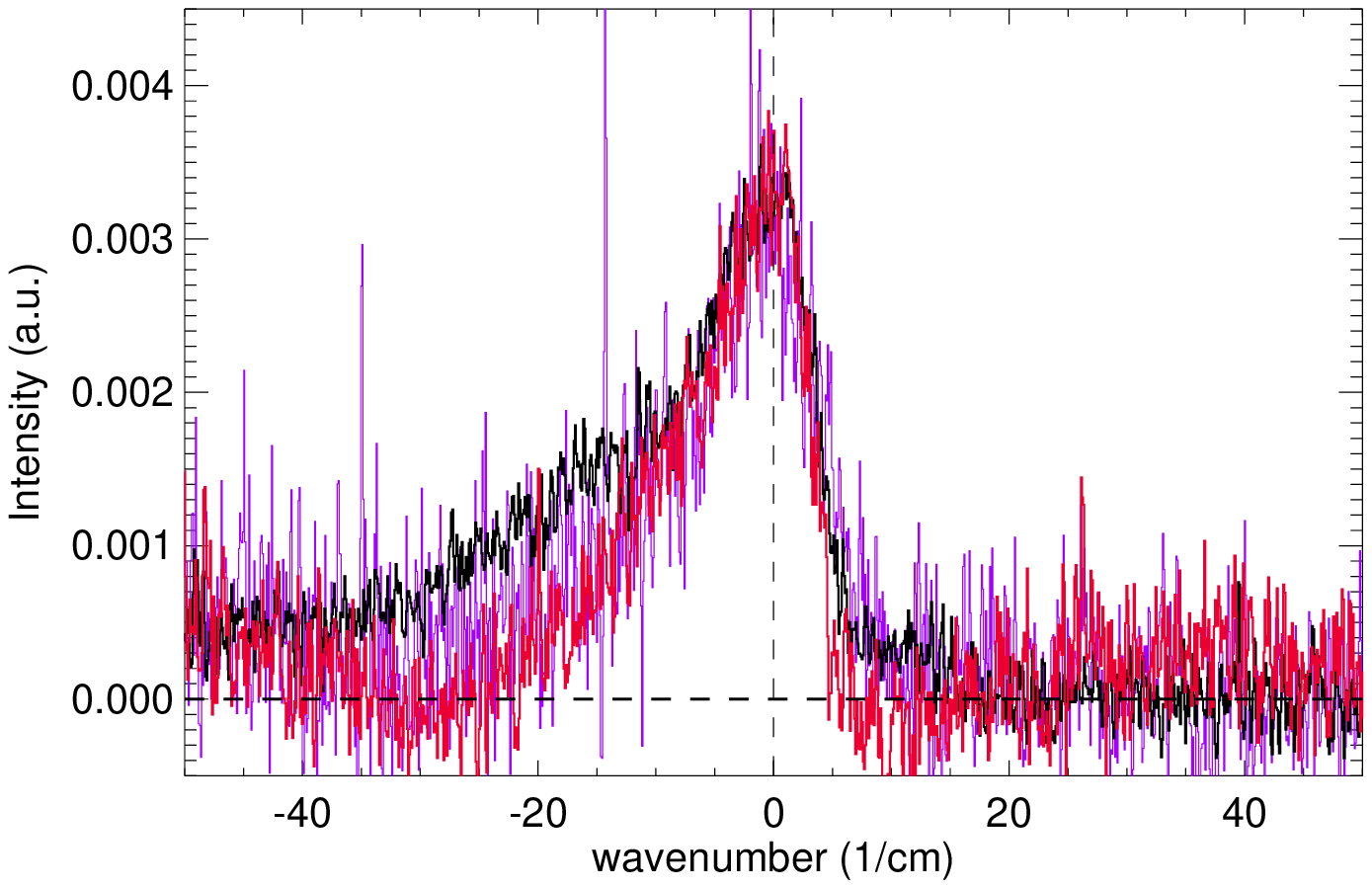}
  \includegraphics[angle=0,width=\columnwidth,clip]{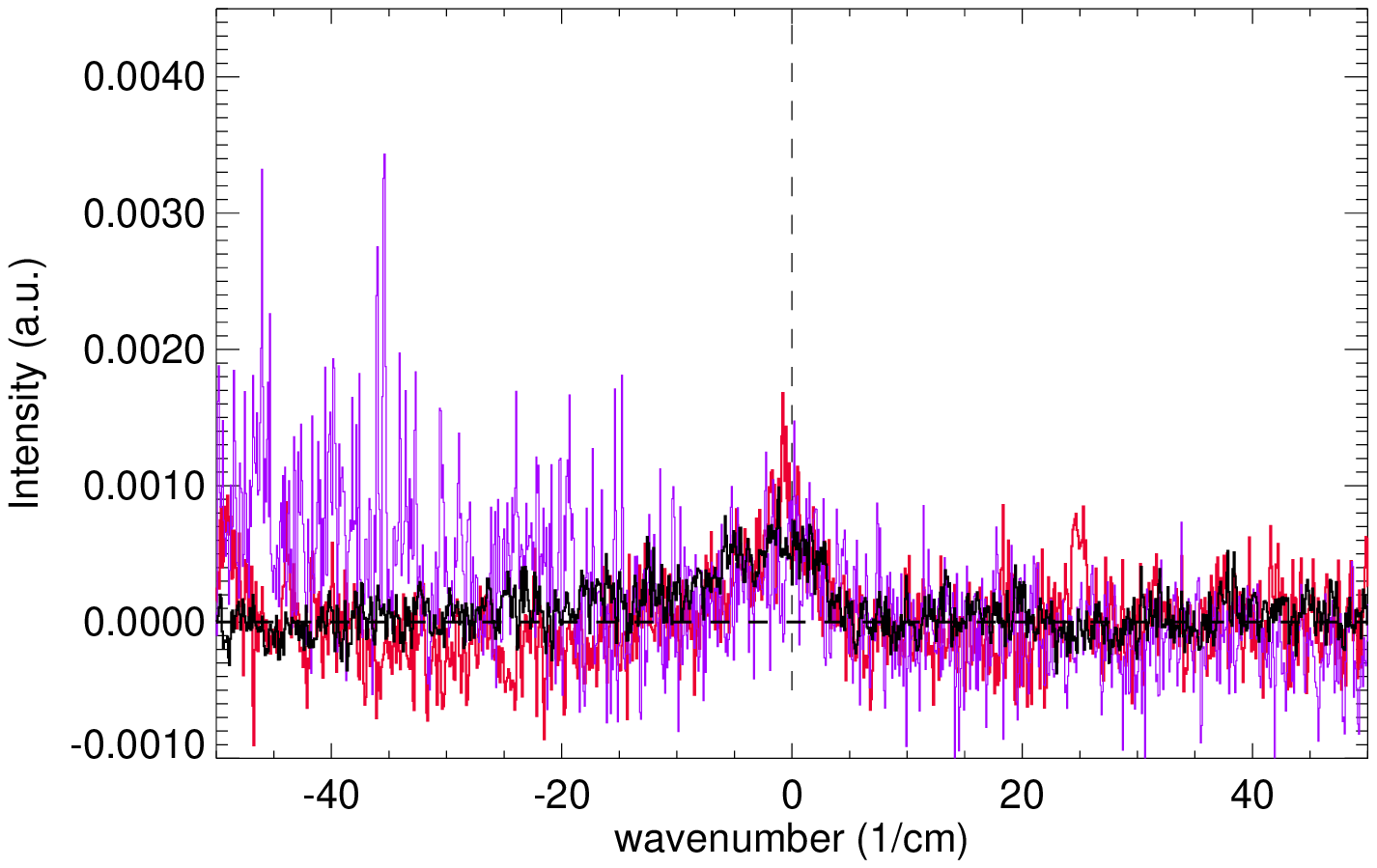}
  \caption{Intensity-wavenumber profiles for the 5800 (black), 5850 (purple) and 6615~\AA\ (red) RRBs 
  compared for the 7\arcsec\ (top) and 13\arcsec\ (bottom) offset positions. 
  The continuum-subtracted profiles have been shifted in wavenumber to optimize the comparison. 
  At 7\arcsec\ the zero wavenumbers for the 3 RRBs correspond to 5798.9, 5852.7, and 6615.4~\AA.
  And at 13\arcsec\ the zero wavenumbers correspond to 5798.3, 5852.3, and 6614.8~\AA, respectively.
  The 5850 and 6615~\AA\ RRB profiles (at both offset positions) have been scaled by factors of 1.75 and 2.5, respectively, 
  to match the peak intensity of the 5800~\AA\ RRB.
  In this figure the displayed spectra have not been smoothed. Both panels are shown on identical  
  wavenumber and intensity ranges. 
  Note also the `inversed' profiles with respect to those plotted as a function of wavelength.}
  \label{fig:RRB5800-5850-6615}
\end{figure}

\begin{figure}[ht!]
 \centering
  \includegraphics[angle=0,width=\columnwidth,clip]{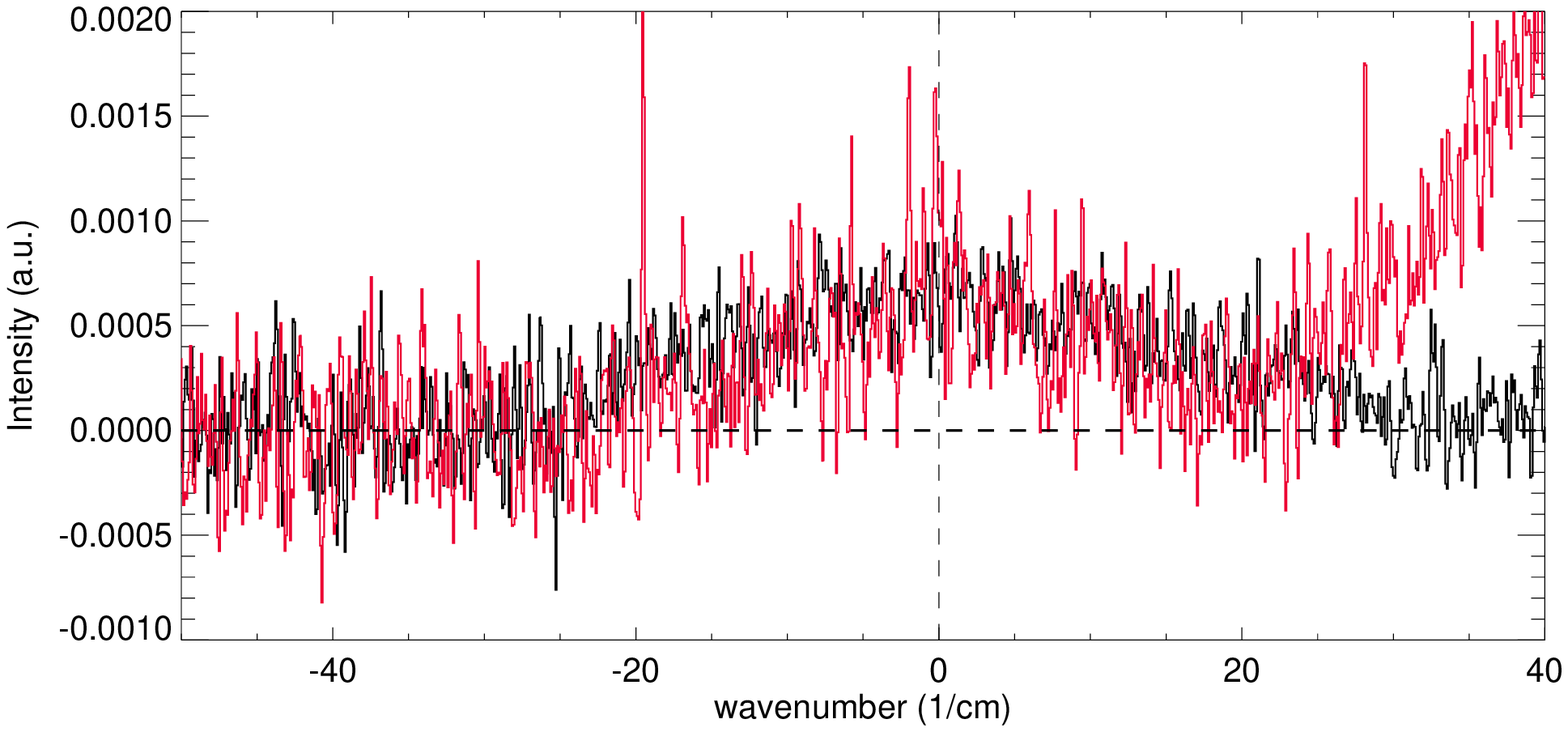}
  \caption{The intensity-wavenumber profiles for the 5825 and 6635~\AA\ RRBs are compared to each other
  for the 7\arcsec\ offset position.
  Each profile has been shifted in wavenumber such as to put the peak at 0~cm$^{-1}$.
  The peak wavelengths correspond to 5826.5 and 6636.5~\AA, respectively.
  The 6635~\AA\ RRB profile has been scaled by a factor of 2.5 to match the peak intensity of the 5825~\AA\ RRB.  
  In this figure the displayed spectra have not been smoothed.}
  \label{fig:RRB5825-6635}
\end{figure}

We note that there exist between individual spectra small differences in strengths and shapes of the nebular emission lines.
These are most likely due to unavoidable (positional) differences between different exposure sequences, and in particular those taken on different nights.
The four spectra taken within a single sequence do not suffer from such shifts as the hole was kept fixed in position with
respect to the central star using Cassegrain tracking.
The total error-weighted average spectra are continuum-subtracted by measuring the continuum levels between 
5780-5790~\AA\ and 6605-6610~\AA, respectively.

The Stokes I, Q and U spectra covering the optical emission bands at 5800~\AA\ (including also the RRB at 5825~\AA) 
and 6615~\AA\ are shown in Figs.~\ref{fig:rrb5} and~\ref{fig:rrb10} for both nebular positions, respectively.
In addition, the check spectra N2(Q) and N2(U) are shown to identify possible spurious polarisation signals.
For clarity the N1 spectra are not show in the above Figures as their appearance is similar to the N2 spectra.
The Null spectra (N1 and N2) are a diagnostic of possible instrumental artifacts linked to the intensity or Stokes parameter, respectively.
Preliminary studies show that for intense emission lines in Stokes I an associated structure in the Null spectra does
not necessarily put in doubt the detection of signatures in the Stokes spectra (\citealt{Fabas2008}).
The central star spectra are displayed at the top in Fig.~\ref{fig:rrb5}. Features are tentatively identified at the position 
of the 5780, 5797 and 6614~\AA\ DIBs.
We identify the three strong features at 5793, 5801, and 6611.5~\AA\ with photospheric \ion{C}{i}, which is known to be present at suprasolar abundances in HD~44179 (\citealt{1992A&A...256L..15W}).

\section{Discussion}

\subsection*{RRB total intensity profiles}
The two RRBs at 5800~\AA\ and 6615~\AA\ are clearly detected at both positions in the nebula (Figs.~\ref{fig:rrb5} and
\ref{fig:rrb10}). The spectra taken at the larger offset show weaker RRBs. 
The intrinsically weaker and broader RRBs at 5826 and 6635~\AA\ are only seen at the position nearest 
to the central star. At 13\arcsec\ a narrow emission component is detected around 6635~\AA.

The emission velocity profiles for the 5800, 5850 and 6615~\AA\ RRBs are very similar for each offset position  as is
illustrated in Fig.~\ref{fig:RRB5800-5850-6615}. Their peak flux ratios are 1:2.5:1.75, respectively.
The FWHM of these RRBs is about 13~cm$^{-1}$ ($\sim$250~\kms) at 7\arcsec\ offset. 
Closer to the star these RRBs have a pronounced red wing which is not present at 13\arcsec\ from the central source.
At 7\arcsec\ the profiles show a steep blue edge as well as a broad red-wing, while
further away from the star, at 13\arcsec\ the wing has disappeared and a weak narrow component (shifted bluewards) remains.
The red wing of the 5800~\AA\ feature appears slightly more enhanced with respect to the two other RRBs.
This behaviour is consistent with the blueshift in the RRB feature and decrease in the FWHM for increasing offsets from the central star 
as discussed by \emph{e.g.} \citet{2002MNRAS.332L..17G} and \citet{2002A&A...390..147V}.

The appearance of a red wing is consistent with the common observation of vibrational sequence bands in electronic spectra 
where the vibrational frequencies in the emitting state are slightly lower than in the ground state. 
With increase in offset, the populations of vibrationally excited levels in the ground electronic state will decline 
resulting in reduced optical excitation to vibrationally excited levels of the excited electronic state. Consequently a 
reduction in sequence band emission intensity with offset would be anti\-ci\-pated and is consistent with narrowing of the profile. 
The blueward edge is realtively unaffected by a reduction in internal vibrational temperature with offset.

In contrast to the 5800, 5850 and 6615 bands, the profiles of the 5825 and 6635~\AA\ RRBs (at 7\arcsec) have symmetrical shapes 
with FWHM of $\sim$24~cm$^{-1}$ (Fig.~\ref{fig:RRB5825-6635}).
These two bands appear to be satellites of the 5800 and 6615 RRBs, respectively, and within this interpretation could arise from transitions from
the excited vibronic (vibrationless) levels to one or more vibrationally excited levels in the ground electronic states of the carriers.  
The separations between the  5800 and 5825~\AA\  (74~cm$^{-1}$), and 6615 and 6635~\AA\  (46~cm$^{-1}$) bands, would then be vibrational 
intervals in the respective ground electronic states of the carriers. 
The band intensities of the two satellite bands are too low in these data to determine their behaviour as a function of offset.  
The fact that they are symmetrical may be significant, although may result simply from the overlap of multiple bands. 

\subsection*{RRB polarisation features}
Inspection of the averaged polarisation spectra at 7\arcsec\ show a small polarisation feature in the 5800~\AA\ RRB for both Stokes Q and U
spectra. However, since these features also appear in the check N1 and N2 spectra they can not be confirmed to be non-artificial 
(\emph{i.e.} due to instrumental effects and not intrinsic to the source).
Preliminary investigations of ESPaDOnS data have shown that intense emission in Stokes I could be associated with a structure in the N channels, 
without putting the reliability of the detected signatures in Stokes parameter into doubt (Fabas 2008). 
This needs to be further investigated.
If these features were real polarisation features intrinsic to the emission profile we note that 
the peak polarisation level of this feature would be $\sim$0.05\%, with a 1-$\sigma$ noise on the polarisation of about $\sim$0.02\%.
None of the other RRBs at 5825, 5850, and 6615~\AA\ show any sign of polarisation features (Figs.~\ref{fig:rrb5} and~\ref{fig:rrb10}).
The absence of line polarisation in the RRBs suggests they are not directly related to dust grains. 
The expected levels of line polarisation would be similar to that of the continuum ($\sim$2\%).
For fluorescent emission due to asymmetric molecules (\emph{i.e.} with a rotational Q-branch) excess line polarisation, up to 20\%, in the Q-branch with respect
to the P- and R-branches is predicted (\citealt{1998JApSp..65..643T}).
We conclude that the RRBs are not likely to be due to grains or molecules with strong Q-branches (such as PAHs with side-groups; 
amino-acids, etc.). Possibly the RRB carriers are more symmetric (thus lacking a Q-branch) due to photo-destruction of side-groups
in the harsh circumstellar UV radiation environment of the Red Rectangle.
Nonetheless, the results are also consistent with the carriers being large gas-phase molecules (even those with a high degree of symmetry)
as statistical effects then play a more significant role in reducing the level of fluorescence polarisation.

\section{Conclusion}
The spectropolarimetric observations presented here give new insight in the properties of the red rectangle bands (RRBs).
The high resolution spectra show that (as suspected from lower resolution spectra) the profiles of the 5800, 5850, and 6615~\AA\ RRBs 
are very similar. Although the profiles change with distance from the central source, this similarity holds for both positions.
These high-resolution spectra reveal no profile substructure except for the appearance of a red wing, as observed before, 
at the smallest offset position.

The main result is that the Red Rectangle Bands at 5800, 5825, 5850, 6615 and 6635~\AA\ are not polarised, down to levels of 0.02\%.
This implies that 
\begin{itemize}
\item the carriers of the RRBs are not directly attached to the dust grains as the latter show significant continuum polarisation;
\item the lack of strong narrow polarisation features further suggest that the Q-branch emission does not make up part of the observed profile.
\item the carriers are large molecules as statistical effects then play a more significant role in reducing the level 
of fluorescence polarisation.
On the other hand, these transitions could be due to a non-vibronic origin band electronic transition, even for a symmetric molecule.
\end{itemize}

\noindent 
Nonetheless, it is known that P- and R-branches giving rise to fluorescent emission can show line polarisation at very low levels. 
Thus, further higher signal-to-noise spectropolarimetry at different positions in the extended emission is needed to 
1) establish the reality of the tentative polarisation feature related to the 5800~\AA\ RRB and 2) to probe in more detail the 
polarisation of all known RRBs as they do not necessarily share identical carriers. 

\begin{acknowledgements}
We thank the QSO staff - in particular Nadine Manset - at the CFHT for advice on preparation for the observing run and on technical aspects of the processing of ESPaDOnS data. 
We thank the referee for helpful comments.
\end{acknowledgements}

\bibliographystyle{aa} 
\bibliography{/lhome/nick/Desktop/ReadingMaterial/Astronomy/Bibtex/bibtex} 

\begin{thebibliography}{52}
\expandafter\ifx\csname natexlab\endcsname\relax\def\natexlab#1{#1}\fi

\bibitem[{{Borisevich} {et~al.}(2002){Borisevich}, {Derzhitskii}, {Povedailo},
  \& {Tolkachev}}]{2002JApSp..69..571B}
{Borisevich}, N.~A., {Derzhitskii}, S.~L., {Povedailo}, V.~A., \& {Tolkachev},
  V.~A. 2002, Journal of Applied Spectroscopy, 69, 571

\bibitem[{{Borisevich} {et~al.}(2006){Borisevich}, {Poveda{\u i}lo},
  {Tolkachev}, \& {Yakovlev}}]{2006DokPh..51...17B}
{Borisevich}, N.~A., {Poveda{\u i}lo}, V.~A., {Tolkachev}, V.~A., \&
  {Yakovlev}, D.~L. 2006, Physics - Doklady, 51, 17

\bibitem[{{Cohen} {et~al.}(1975){Cohen}, {Anderson}, {Cowley}, {Coyne},
  {Fawley}, {Gull}, {Harlan}, {Herbig}, {Holden}, {Hudson}, {Jakoubek},
  {Johnson}, {Merrill}, {Schiffer}, {Soifer}, \&
  {Zuckerman}}]{1975ApJ...196..179C}
{Cohen}, M., {Anderson}, C.~M., {Cowley}, A., {et~al.} 1975, \apj, 196, 179

\bibitem[{{Cohen} {et~al.}(2004){Cohen}, {Van Winckel}, {Bond}, \&
  {Gull}}]{2004AJ....127.2362C}
{Cohen}, M., {Van Winckel}, H., {Bond}, H.~E., \& {Gull}, T.~R. 2004, \aj, 127,
  2362

\bibitem[{{Darbon} {et~al.}(2000){Darbon}, {Zavagno}, {Perrin}, {Savine},
  {Ducci}, \& {Sivan}}]{2000A&A...364..723D}
{Darbon}, S., {Zavagno}, A., {Perrin}, J., {et~al.} 2000, \aap, 364, 723

\bibitem[{{d'Hendecourt} {et~al.}(1986){d'Hendecourt}, {L{\'e}ger}, {Olofsson},
  \& {Schmidt}}]{1986A&A...170...91D}
{d'Hendecourt}, L.~B., {L{\'e}ger}, A., {Olofsson}, G., \& {Schmidt}, W. 1986,
  \aap, 170, 91

\bibitem[{{Donati} {et~al.}(1997){Donati}, {Semel}, {Carter}, {Rees}, \&
  {Collier Cameron}}]{1997MNRAS.291..658D}
{Donati}, J.-F., {Semel}, M., {Carter}, B.~D., {Rees}, D.~E., \& {Collier
  Cameron}, A. 1997, \mnras, 291, 658

\bibitem[{{Fabas}(2008)}]{Fabas2008}
{Fabas}, N. 2008, Master thesis, Lulea University of Technology

\bibitem[{{Fossey}(1991)}]{1991Natur.353..393F}
{Fossey}, S.~J. 1991, \nat, 353, 393

\bibitem[{{Furton} \& {Witt}(1990)}]{1990ApJ...364L..45F}
{Furton}, D.~G. \& {Witt}, A.~N. 1990, \apjl, 364, L45

\bibitem[{{Geballe} {et~al.}(1989){Geballe}, {Tielens}, {Allamandola},
  {Moorhouse}, \& {Brand}}]{1989ApJ...341..278G}
{Geballe}, T.~R., {Tielens}, A.~G.~G.~M., {Allamandola}, L.~J., {Moorhouse},
  A., \& {Brand}, P.~W.~J.~L. 1989, \apj, 341, 278

\bibitem[{{Gledhill} {et~al.}(2009){Gledhill}, {Witt}, {Vijh}, \&
  {Davis}}]{2009MNRAS.392.1217G}
{Gledhill}, T.~M., {Witt}, A.~N., {Vijh}, U.~P., \& {Davis}, C.~J. 2009,
  \mnras, 392, 1217

\bibitem[{{Glinski} \& {Anderson}(2002)}]{2002MNRAS.332L..17G}
{Glinski}, R.~J. \& {Anderson}, C.~M. 2002, \mnras, 332, L17

\bibitem[{{Glinski} {et~al.}(2009){Glinski}, {Michaels}, {Anderson}, {Schmidt},
  {Sharp}, {Sitko}, {Bernstein}, \& {van Winckel}}]{2009Ap&SS.323..337G}
{Glinski}, R.~J., {Michaels}, P.~D., {Anderson}, C.~M., {et~al.} 2009, \apss,
  323, 337

\bibitem[{{Gordon} {et~al.}(1998){Gordon}, {Witt}, \&
  {Friedmann}}]{1998ApJ...498..522G}
{Gordon}, K.~D., {Witt}, A.~N., \& {Friedmann}, B.~C. 1998, \apj, 498, 522

\bibitem[{{Greenstein} \& {Oke}(1977)}]{1977PASP...89..131G}
{Greenstein}, J.~L. \& {Oke}, J.~B. 1977, \pasp, 89, 131

\bibitem[{{Hony} {et~al.}(2001){Hony}, {Van Kerckhoven}, {Peeters}, {Tielens},
  {Hudgins}, \& {Allamandola}}]{2001A&A...370.1030H}
{Hony}, S., {Van Kerckhoven}, C., {Peeters}, E., {et~al.} 2001, \aap, 370, 1030

\bibitem[{{Merrill}(1977)}]{1977ivsw.conf..446M}
{Merrill}, K.~M. 1977, in IAU Colloq. 42: The Interaction of Variable Stars
  with their Environment, ed. {R.~Kippenhahn, J.~Rahe, \& W.~Strohmeier}, 446

\bibitem[{{Nathanson} \& {McClelland}(1986)}]{1986JChPh..85.4311N}
{Nathanson}, G.~M. \& {McClelland}, G.~M. 1986, \jcp, 85, 4311

\bibitem[{{Peeters} {et~al.}(2002){Peeters}, {Hony}, {Van Kerckhoven},
  {Tielens}, {Allamandola}, {Hudgins}, \& {Bauschlicher}}]{2002A&A...390.1089P}
{Peeters}, E., {Hony}, S., {Van Kerckhoven}, C., {et~al.} 2002, \aap, 390, 1089

\bibitem[{{Perkins} {et~al.}(1981){Perkins}, {Scarrott}, {Murdin}, \&
  {Bingham}}]{1981MNRAS.196..635P}
{Perkins}, H.~G., {Scarrott}, S.~M., {Murdin}, P., \& {Bingham}, R.~G. 1981,
  \mnras, 196, 635

\bibitem[{{Reese} \& {Sitko}(1996)}]{1996ApJ...467L.105R}
{Reese}, M.~D. \& {Sitko}, M.~L. 1996, \apjl, 467, L105

\bibitem[{{Russell} {et~al.}(1977){Russell}, {Soifer}, \&
  {Merrill}}]{1977ApJ...213...66R}
{Russell}, R.~W., {Soifer}, B.~T., \& {Merrill}, K.~M. 1977, \apj, 213, 66

\bibitem[{{Russell} {et~al.}(1978){Russell}, {Soifer}, \&
  {Willner}}]{1978ApJ...220..568R}
{Russell}, R.~W., {Soifer}, B.~T., \& {Willner}, S.~P. 1978, \apj, 220, 568

\bibitem[{{Sarre}(1991)}]{1991Natur.351..356S}
{Sarre}, P.~J. 1991, \nat, 351, 356

\bibitem[{{Sarre}(2006)}]{2006JMoSp.238....1S}
{Sarre}, P.~J. 2006, Journal of Molecular Spectroscopy, 238, 1

\bibitem[{{Sarre} {et~al.}(1995){Sarre}, {Miles}, \&
  {Scarrott}}]{1995Sci...269..674S}
{Sarre}, P.~J., {Miles}, J.~R., \& {Scarrott}, S.~M. 1995, Science, 269, 674

\bibitem[{{Scarrott} {et~al.}(1992){Scarrott}, {Watkin}, {Miles}, \&
  {Sarre}}]{1992MNRAS.255P..11S}
{Scarrott}, S.~M., {Watkin}, S., {Miles}, J.~R., \& {Sarre}, P.~J. 1992,
  \mnras, 255, 11P

\bibitem[{{Schmidt} {et~al.}(1980){Schmidt}, {Cohen}, \&
  {Margon}}]{1980ApJ...239L.133S}
{Schmidt}, G.~D., {Cohen}, M., \& {Margon}, B. 1980, \apjl, 239, L133

\bibitem[{{Schmidt} \& {Witt}(1991)}]{1991ApJ...383..698S}
{Schmidt}, G.~D. \& {Witt}, A.~N. 1991, \apj, 383, 698

\bibitem[{{Sharp} {et~al.}(2006){Sharp}, {Reilly}, {Kable}, \&
  {Schmidt}}]{2006ApJ...639..194S}
{Sharp}, R.~G., {Reilly}, N.~J., {Kable}, S.~H., \& {Schmidt}, T.~W. 2006,
  \apj, 639, 194

\bibitem[{{Sironi} \& {Draine}(2009)}]{2009ApJ...698.1292S}
{Sironi}, L. \& {Draine}, B.~T. 2009, \apj, 698, 1292

\bibitem[{{Sloan} {et~al.}(1993){Sloan}, {Grasdalen}, \&
  {Levan}}]{1993ApJ...409..412S}
{Sloan}, G.~C., {Grasdalen}, G.~L., \& {Levan}, P.~D. 1993, \apj, 409, 412

\bibitem[{{Song} {et~al.}(2007){Song}, {McCombie}, {Kerr}, \&
  {Sarre}}]{2007MNRAS.380..979S}
{Song}, I., {McCombie}, J., {Kerr}, T.~H., \& {Sarre}, P.~J. 2007, \mnras, 380,
  979

\bibitem[{{Tielens}(2008)}]{2008ARA&A..46..289T}
{Tielens}, A.~G.~G.~M. 2008, \araa, 46, 289

\bibitem[{{Tolkachev}(1998{\natexlab{a}})}]{1998JApSp..65..883T}
{Tolkachev}, V.~A. 1998{\natexlab{a}}, Journal of Applied Spectroscopy, 65, 883

\bibitem[{{Tolkachev}(1998{\natexlab{b}})}]{1998JApSp..65..643T}
{Tolkachev}, V.~A. 1998{\natexlab{b}}, Journal of Applied Spectroscopy, 65, 643

\bibitem[{{Tolkachev}(2002)}]{2002JApSp..69..711T}
{Tolkachev}, V.~A. 2002, Journal of Applied Spectroscopy, 69, 711

\bibitem[{{Tolkachev} \& {Blokhin}(2009)}]{2009JApSp..76..806T}
{Tolkachev}, V.~A. \& {Blokhin}, A.~P. 2009, Journal of Applied Spectroscopy,
  76, 806

\bibitem[{{Tolkachev} \& {Pliska}(1986)}]{1986OptSp..61...48T}
{Tolkachev}, V.~A. \& {Pliska}, S.~P. 1986, Optics and Spectroscopy, 61, 48

\bibitem[{{Tolkachev} \& {Polubisok}(1995)}]{1995SpecL..28..441T}
{Tolkachev}, V.~A. \& {Polubisok}, S.~A. 1995, Spectroscopy Letters, 28, 441

\bibitem[{{Tolkachev} \& {Polubisok}(1998)}]{1998OptSp..84..839T}
{Tolkachev}, V.~A. \& {Polubisok}, S.~A. 1998, Optics and Spectroscopy, 84, 839

\bibitem[{{Van Winckel} {et~al.}(2002){Van Winckel}, {Cohen}, \&
  {Gull}}]{2002A&A...390..147V}
{Van Winckel}, H., {Cohen}, M., \& {Gull}, T.~R. 2002, \aap, 390, 147

\bibitem[{{Vijh} {et~al.}(2004){Vijh}, {Witt}, \&
  {Gordon}}]{2004ApJ...606L..65V}
{Vijh}, U.~P., {Witt}, A.~N., \& {Gordon}, K.~D. 2004, \apjl, 606, L65

\bibitem[{{Vijh} {et~al.}(2005){Vijh}, {Witt}, \&
  {Gordon}}]{2005ApJ...633..262V}
{Vijh}, U.~P., {Witt}, A.~N., \& {Gordon}, K.~D. 2005, \apj, 633, 262

\bibitem[{{Wada} {et~al.}(2009){Wada}, {Mizutani}, {Narisawa}, \&
  {Tokunaga}}]{2009ApJ...690..111W}
{Wada}, S., {Mizutani}, Y., {Narisawa}, T., \& {Tokunaga}, A.~T. 2009, \apj,
  690, 111

\bibitem[{{Waelkens} {et~al.}(1992){Waelkens}, {Van Winckel}, {Trams}, \&
  {Waters}}]{1992A&A...256L..15W}
{Waelkens}, C., {Van Winckel}, H., {Trams}, N.~R., \& {Waters}, L.~B.~F.~M.
  1992, \aap, 256, L15

\bibitem[{{Waelkens} {et~al.}(1996){Waelkens}, {Waters}, {de Graauw}, {Huygen},
  {Malfait}, {Plets}, {Vandenbussche}, {Beintema}, {Boxhoorn}, {Habing},
  {Heras}, {Kester}, {Lahuis}, {Morris}, {Roelfsema}, {Salama}, {Siebenmorgen},
  {Trams}, {van der Bliek}, {Valentijn}, \& {Wesselius}}]{1996A&A...315L.245W}
{Waelkens}, C., {Waters}, L.~B.~F.~M., {de Graauw}, M.~S., {et~al.} 1996, \aap,
  315, L245

\bibitem[{{Warren-Smith} {et~al.}(1981){Warren-Smith}, {Scarrott}, \&
  {Murdin}}]{1981Natur.292..317W}
{Warren-Smith}, R.~F., {Scarrott}, S.~M., \& {Murdin}, P. 1981, \nat, 292, 317

\bibitem[{{Waters} {et~al.}(1998){Waters}, {Waelkens}, {Van Winckel},
  {Molster}, {Tielens}, {van Loon}, {Morris}, {Cami}, {Bouwman}, {de Koter},
  {de Jong}, \& {de Graauw}}]{1998Natur.391..868W}
{Waters}, L.~B.~F.~M., {Waelkens}, C., {Van Winckel}, H., {et~al.} 1998, \nat,
  391, 868

\bibitem[{{Witt} \& {Boroson}(1990)}]{1990ApJ...355..182W}
{Witt}, A.~N. \& {Boroson}, T.~A. 1990, \apj, 355, 182

\bibitem[{{Witt} {et~al.}(2006){Witt}, {Gordon}, {Vijh}, {Sell}, {Smith}, \&
  {Xie}}]{2006ApJ...636..303W}
{Witt}, A.~N., {Gordon}, K.~D., {Vijh}, U.~P., {et~al.} 2006, \apj, 636, 303

\end{thebibliography}

\end{document}